\documentclass[final]{svjour3}
\usepackage[dvipdfmx]{graphicx}
\usepackage{rotating}
\usepackage{amssymb}
\usepackage{mathptmx}
\usepackage[numbers,sort&compress]{natbib}
\usepackage{xcolor}
\usepackage{caption}
\usepackage{subcaption}
\makeatletter
\journalname{Journal of Low Temperature Physics}
\bibpunct{[}{]}{,}{n}{}{,}

\begin{document}

\newcommand{\hdblarrow}{H\makebox[0.9ex][l]{$\downdownarrows$}-}
\title{The Simulation and Design of an On-Chip Superconducting Millimetre Filter-Bank Spectrometer}

\author{G.~Robson$^1$ \and A.~J.~Anderson$^3$ \and P.~S.~Barry$^{1,2,4}$ \and S.~Doyle$^1$ \and K.~S.~Karkare$^{2,3}$ 
    }

\institute{\email{RobsonG2@Cardiff.ac.uk}\\
1: School of Physics \& Astronomy, Cardiff University, Cardiff CF24 3AA, UK\\
2: Kavli Institute for Cosmological Physics, University of Chicago, Chicago, IL 60637, USA\\
3: Fermi National Accelerator Laboratory, Batavia, IL 60510, USA\\
4: High-Energy Physics Division, Argonne National Laboratory, Lemont, IL 60439, USA\\
}

\maketitle

\begin{abstract}
Superconducting on-chip filter-banks provide a scalable, space saving solution to create imaging spectrometers at millimetre and sub-millimetre wavelengths. We present an easy to realise, lithographed superconducting filter design with a high tolerance to fabrication error. Using a capacitively coupled $\lambda/2$ microstrip resonator to define a narrow ($\lambda/\Delta\lambda = 300$) spectral pass band, the filtered output of a given spectrometer channel directly connects to a lumped element kinetic inductance detector. We show the tolerance analysis of our design, demonstrating $<11\%$ change in filter quality factor to any one realistic fabrication error and a full filter-bank efficiency forecast to be 50\% after accounting for fabrication errors and dielectric loss tangent.


\end{abstract}

\section{Introduction}
    \label{sec:Intro}

Millimetre wavelength astronomy contains a wealth of largely untapped information about the universe, ranging from measurements of the cosmic microwave background, to high redshift sources, to the inner workings of the dust-enshrouded regions of space. In the millimetre range, current, conventional spectrometer technology cannot scale up effectively to meet the sensitivity requirements for next-generation science cases. However, by capitalising on the small energy gap of superconductors, high sensitivity, and ease of microwave circuit integration, superconducting filter-bank spectrometers (FBS) will be central in a range of future surveys in the millimetre region of the spectrum.\newline
\indent FBS make use of the low-loss properties of superconductors operating well below the superconducting critical temperature ($T_c$), integrated microwave circuitry, and the sensitive and multiplexing capabilities of kinetic inductance detectors (KIDs) to create an efficient, compact, and sensitive spectrometer that does not need large dispersive optical components, complex readout systems, or noise limiting amplifiers and local oscillators.\newline
\indent This paper presents a novel filter geometry with the application tailored to a medium-resolution mm-wave FBS, which to date has been demonstrated in two alternative instruments \cite{SuperSpec2015, Deshima2019}. This FBS is motivated by the SPT-SLIM project which is discussed in more detail in \cite{KarkareLTD2021, BarryLTD2021}. Each filter-bank will consist of 200 filter channels with a spectral resolution of $R=300$ covering the $2\,$mm atmospheric transmission window ($120\,$GHz - $180\,$GHz).

\section{On-Chip Filter-Bank Spectrometer Design}
    \label{sec:SpectrometerDesign}
  
%The basic operation for these on-chip spectrometer devices is that the incident light couples to a microstrip feedline via an antenna. The coupled light travels along this which is coupled to a series of parallel half wave ($\lambda/2$) resonators to create various spectral channels, each with a different resonant frequency, $f_0$ and 3dB bandwidth, $\Delta f$.

Incident light is coupled onto a transmission line via an antenna and guided to the filter-bank. The filter-bank is made up of a series of capacitively-coupled half-wave ($\lambda/2$) resonators each with a different resonant frequency, $f_0$, set by the length, and spectral bandwidth, $\Delta f$, set by the coupling to the feedline.

\indent The two key performance metrics of an on-chip spectrometer are the spectral resolution and the filter efficiency. The resolution, $R$, is equivalent to the quality factor, $Q_\textrm{filt}$, of each filter channel's resonator,% which is proportional to the ratio of energy stored in the electromagnetic fields to that lost per second in the resonator circuit,
\begin{equation}
    \label{eq:resolutionEquation}
    R = Q_\textrm{filt} = \frac{f_0}{\Delta f} %= \omega \frac{\textrm{Energy Stored}}{\textrm{Energy Loss/Second}}.
\end{equation}

\indent The filter efficiency is defined as the ratio of the total signal output through the channels to that coupled to the FBS. The maximum power transferred through the filter to the detector that can be achieved for a feedline-filter T-junction such as this is 50\% of the incident signal. Note that this is only the limit for a single filter; the overall FBS efficiency can exceed this with over-sampled filters \cite{kov_cs_2012}.

\section{Filter Geometry}
    \label{sec:Geometry}
    
\begin{figure}
        \centering
        \includegraphics[scale= 0.5]{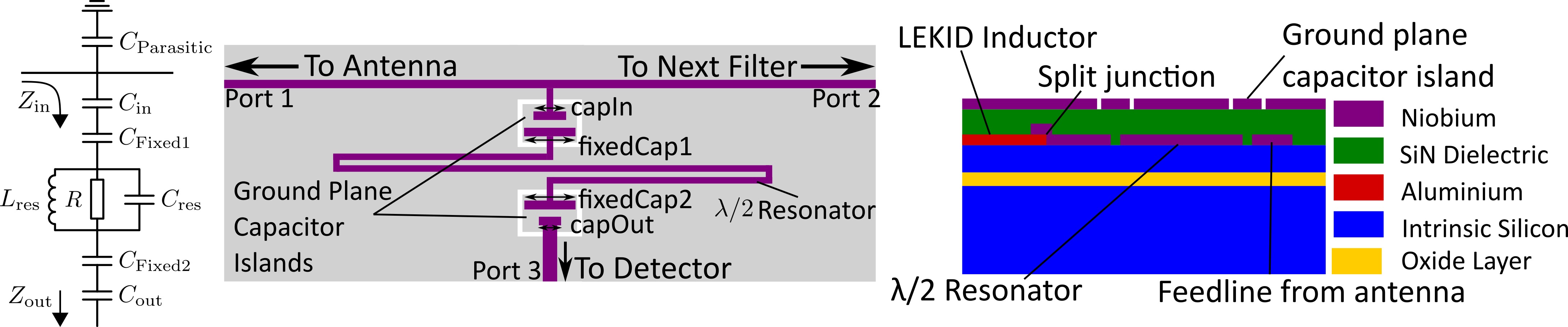}
        \caption{(Color online) (\textit{Left}) lumped-element circuit diagram of the filter where the $\lambda/2$ resonator has been modelled as an LRC circuit with the resistor representing the loss mechanisms. (\textit{Center}) Top-down view of a filter channel microstrip geometry. (\textit{Right}) Schematic of the SOI wafer structure in the region of a filter channel.}
        \label{fig:WaferStructure&Geometry}
\end{figure}

The device will be built from a silicon on oxide (SOI) substrate (see Fig. \ref{fig:WaferStructure&Geometry}) due to requirements for the architecture of the orthomode transducer (OMT) antennas and detectors. For further details see \cite{BarryLTD2021}. Niobium is used for the feedline and filter structure since it is nearly lossless for the range of operation, due to its large superconducting gap.

\indent A filter channel is created by coupling a $\lambda/2$ resonator to the main filter-bank feedline. On resonance, the current distribution results in a current node at each end of the resonator. This enables two coupling methods: inductively, by exploiting the current maximum in the centre as demonstrated by SuperSpec \cite{SuperSpec2015,Shirokoff2012}; or capacitively, using the voltage maximum at the ends, as seen in the DESHIMA \cite{Deshima2019,Endo2013} device. Both examples tune the coupling quality factors, $Q_C$, via the separation between feedline and resonator. Due to fabrication tolerances, the tuning precision of $Q_C$ is somewhat limited for such proximity couplings. We present an alternative coupling scheme that uses a pseudo-lumped element capacitor island in the ground plane to form parallel plate capacitors; two at either end of the resonator, as shown in Fig. \ref{fig:WaferStructure&Geometry}. This allows for a highly tuneable capacitive coupling using a microstrip architecture to minimize losses due to radiation \cite{DelftSpectrometer2019}.

\indent The spectral resolution of a single channel can be written as the inverse sum of the quality factors due to losses within resonator, $Q_\textrm{loss}$, the coupling to the feedline, $Q_\textrm{feed}$, and detector coupling, $Q_\textrm{det}$, \cite{SuperSpec2015},

\begin{equation}
    \label{eq:FilterResolution}
    \frac{1}{R} = \frac{1}{Q_\textrm{filt}} =  \frac{1}{Q_\textrm{C}} +  \frac{1}{Q_\textrm{loss}} = \frac{1}{Q_\textrm{feed}} +\frac{1}{Q_\textrm{det}} + \frac{1}{Q_\textrm{loss}}.
\end{equation}

$Q_\textrm{feed}$ and $Q_\textrm{det}$ can be controlled by changing the overlapping surface area of the parallel plates. For this design,  $Q_\textrm{filt}$ is largely controlled by the capacitor widths ``capIn'' and ``capOut'' (see Fig. and $f_0$ by the resonator length (see Fig. \ref{fig:WaferStructure&Geometry}). Using Sonnet EM \cite{Sonnet} to simulate the geometry in Fig.~\ref{fig:WaferStructure&Geometry}, for a variety of resonator lengths across the band, a simulation sweeping capIn values was carried out with the port 3 impedance set high enough to resemble an open circuit. Fitting Lorentzian curves, the capIn yielding a $Q_\textrm{feed}=600$ can be interpolated. This value is then used whilst sweeping capOut with a matched load at port 3 to obtain the capOut value giving a peak value of 0.5 in the $|\textrm{S}_{31}|^2$. This should ideally be when $Q_\textrm{C}=300$, yet this is not quite the case since the inability to fully de-embed the port 3 output line limits the complete removal of the contribution to $Q_\textrm{C}$ of capOut when optimising capIn. This data was used to build an interpolation table that yields the required geometry needed for a filter with any desired f0.
\newline  There is a small section of microstrip between the feedline and ``capIn'', which, off-resonance acts as a very short stub, which is well modelled as a parasitic capacitance to ground. An effect analogous to the impedance engineering in superconducting parametric amplifiers \cite{Eom2012} occurs when multiple small mismatches are present along the feedline. From our simulations, to prevent the resulting stopband entering into the operational frequency band, the electrical length between filters must be suitably small ($<\lambda/4$) compared to the smallest half wavelength in the band.

\section{Loss Tangent Impact}
    \label{sec:LossTangentImpact}
    
The presence of a dielectric layer introduces a loss mechanism due to parasitic two level systems (TLS), which has been extensively studied \cite{GoaThesis2008, Pappas2011}. The dielectric loss tangent, $\textrm{tan}\delta$, is a measure of the energy dissipated through this mechanism and hence can be expressed as a quality factor and is considered to be the dominant contribution to $Q_\textrm{loss}$ in Eq.  \ref{eq:FilterResolution} thus, $\textrm{tan}\delta = 1/Q_\textrm{loss}$.
Whilst there are limited published results for the loss tangent of SiN, those that are found in the literature are measured at microwave frequencies below $10\,\textrm{GHz}$ \cite{Martinis2005, PaikHanhee2010Rqdl} and require a rough extrapolation of the frequency dependence up to the relevant frequencies for mm-wave astronomy. Moreover, the dielectric tan$\delta$ varies considerably depending on the deposition system and process, thus it will be crucial to characterise our SiN tan$\delta$. For this study, we assume a  SiN dielectric tan$\delta$ of $7\times10^{-4}$~\cite{Hailey-DunsheathS2014OMoS}, since this is similar in process and device architecture.

When optimised, the maximal fraction of power, $\eta$ that passes through the $\lambda/2$ filter and terminates at the detector (filter efficiency) depends on the ratio of $Q_\textrm{filt}$ to $Q_\textrm{loss}$ as \cite{Shirokoff2012}

        \begin{equation}
            \label{eq:MaxFilterEfficiency}
            \eta_\textrm{max} = \frac{1}{2}\left[1 - \frac{Q_\textrm{filt}}{Q_\textrm{loss}} \right]^2.
        \end{equation}

\begin{figure}[t]
\centering
    \includegraphics[scale=0.27]{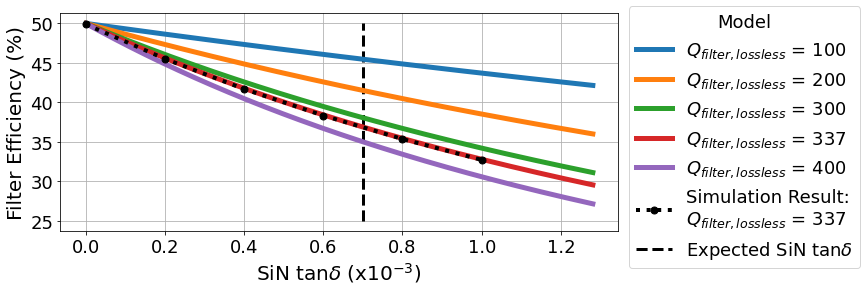}
    \caption{(Color online) The trade off between the filter resolution, $R$ and efficiency, $\eta$ for a range of initial lossless filter quality factors, $Q_\textrm{filt, lossless}$ and tan$\delta$ values.  Note, to compare to the simulation, the model $\textrm{tan}\delta$ has been scaled by the effective dielectric constant of the inverted microstrip geometry.}
    
    \label{fig:LossTangentStudy}
\end{figure}

Fig. \ref{fig:LossTangentStudy} shows the impact on $\eta$ that the loss tangent has over a range of $Q_\textrm{filt, lossless}$ predicted by Eq. \ref{eq:MaxFilterEfficiency}, as well as a Sonnet simulation corrected for the effective dielectric constant, which shows good agreement with a $Q_\textrm{filt, lossless} = 337$ filter at $148\,$GHz. There is also the expected trade off between $R$ and $\eta$ for a given dielectric tan$\delta$, and Fig.~\ref{fig:LossTangentStudy} clearly show what stands to be gained from lower loss dielectrics such as $1\times 10^{-4}$  as is expected from amorphous silicon \cite{Deshima2019}.

\section{Fabrication Error Tolerance}
    \label{sec:FabErrorTolerance}

\begin{figure}[t]
    \centering
    \includegraphics[scale=0.3]{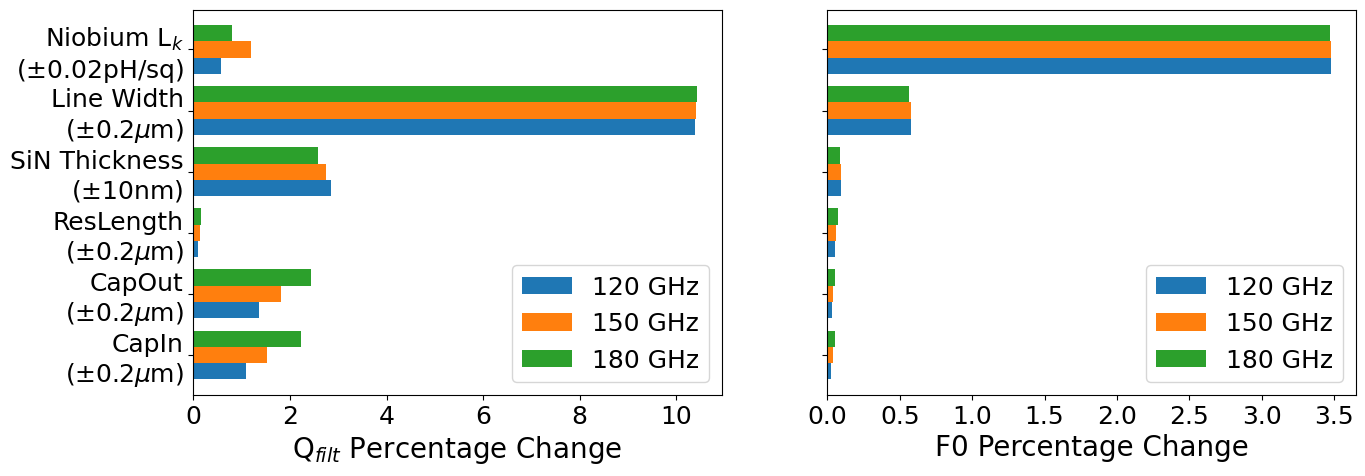}
    \caption{(Color online) Simulated single-filter quality factor, $Q_{filt}$, (\textit{Left}) and resonant frequency, $f_0$, (\textit{Right}) tolerance to the worst-case fabrication errors for different variables.}
    \label{fig:QfiltF0FabricationTolerance}
\end{figure}

The largest source for disagreement between simulated and fabricated filters is generally due to fabrication error, a somewhat unavoidable aspect of micro-fabrication. Deviations in the kinetic inductance, $L_k$, of the superconducting film, the dielectric thickness/constant, and over or under etching, have all been shown to have a considerable impact on the filter properties, and this is particularly the case for etch sensitivity with proximity-coupled filters \cite{Laguna2021}. Therefore it is necessary to understand the tolerance of a design to a variety of fabrication errors. Fig.~\ref{fig:QfiltF0FabricationTolerance} shows the filters' $Q_\textrm{filt, lossless}$ and $f_0$ sensitivity to variations in key features of our architecture using the python model described in Sec. \ref{sec:FilterBankPerformance}, to study the impact of typical deposition and lithography error based on unoptimised fabrication runs at Cardiff.\newline
\indent These results clearly demonstrate that this design has a high tolerance to fabrication error, particularly given that worst case errors were used. The resonant frequency is most sensitive to variations in $L_k$, however this is only approximately $3.5\,\%$ and suitably constant across the band. Thus, all filters should shift by the same amount. Note, changes in $L_k$ due to a variation in niobium thickness were not considered as $L_k$ for niobium is expected to vary slowly with thickness for $\approx 300$~nm thick niobium with a magnetic penetration depth of $\approx135$ nm. Furthermore, we see a promising forecast for the tolerance of the geometry to the variables that are the main controllers of $Q_C$, namely ``CapIn'', ``CapOut'', and ``SiN Thickness'' since these control the capacitance of the coupling capacitors, each resulting in a shift in Q between 2-3\%. By comparison, proximity-coupled filters demonstrate a far higher sensitivity to etch error at the coupling geometry where a change in $100\,\textrm{nm}$ can triple the value of $Q_C$~\cite{Laguna2021}. This low tolerance removes the need to use electron beam lithography, a more precise but time-consuming etching method, reducing the cost, and speed of device turn around.

\section{Filter-Bank Performance}
    \label{sec:FilterBankPerformance}

\begin{figure}
     \centering
     \begin{subfigure}[b]{0.42\textwidth}
         \centering
         \includegraphics[width=\textwidth]{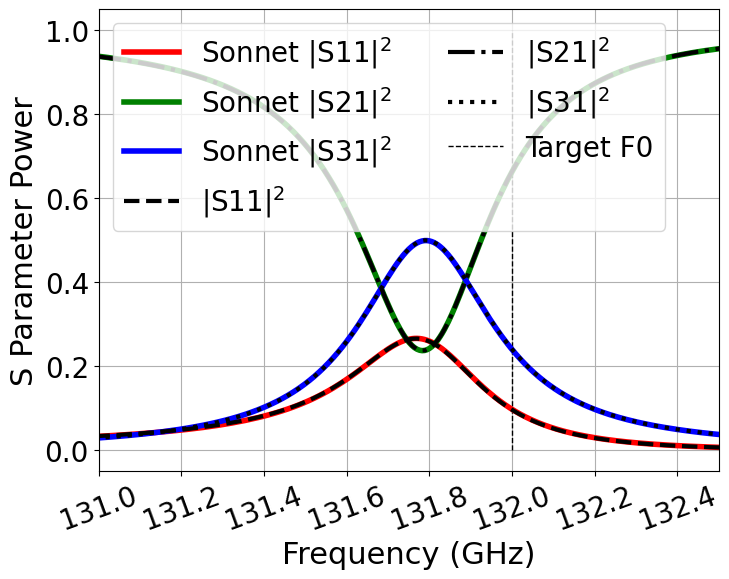}
     \end{subfigure}
     \hfill
     \begin{subfigure}[b]{0.458\textwidth}
         \centering
         \includegraphics[width=\textwidth]{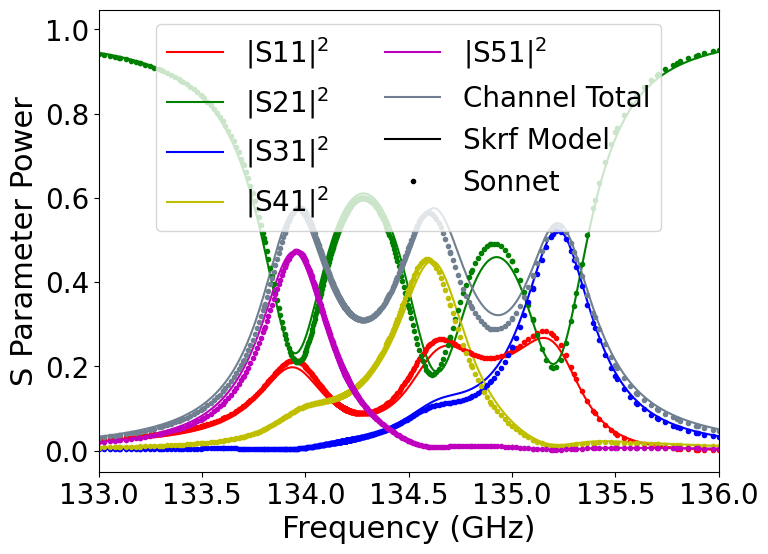}
     \end{subfigure}
        \caption{(Color online) Comparison of the S-Parameters simulated by the Scikit-rf python model to that output by Sonnet for a single filter with an intended $f_0$ of $132\,$GHz (\textit{Left}) and a three filter filter-bank with $f_0 = 134.0, 134.5 \textrm{ and } 135.0$ (\textit{Right}).}
        \label{fig:SkrfModelComparisons}
\end{figure}

\begin{figure}
    \centering
    \includegraphics[scale=0.41]{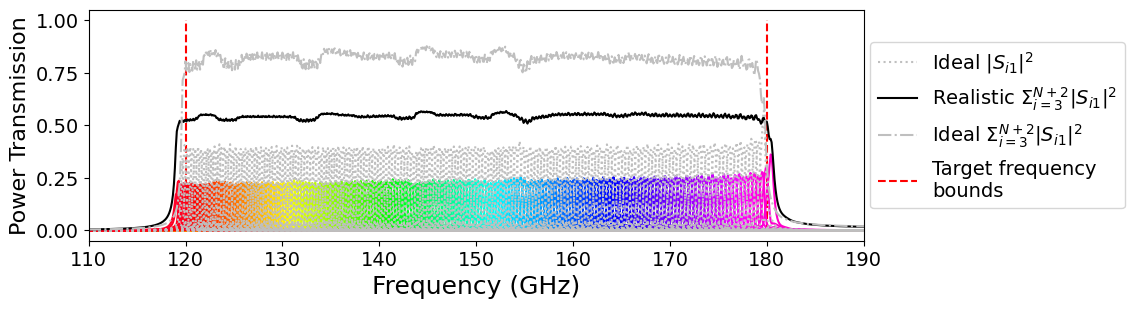}
    \caption{(Color online) Two python model simulations of an SPT-SLIM filter bank between $120 - 180\,$GHz with an oversampling, $\Sigma = 1.6$ showing the single and total channel power transmission for an ideal and a realistic FBS. The colored profiles are realistic channel responses. The ``Realistic'' simulation includes etch error, tan$\delta = 7\times10^{-4}$ and SiN thickness variation profiles over the length of the FBS based on data from unoptimised fabrication runs at Cardiff. Whereas the ``Ideal'' results do not include fabrication error nor loss tangent.}
    \label{fig:filterbank120-180GHzFullFabError}
\end{figure}

Sonnet EM captures electromagnetic behaviour otherwise overlooked by lumped-element simulation software. This is at the cost of simulation run time, and therefore unsuitable for simulating the performance of an FBS with more than three filters. Furthermore, niche issues start to become a problem such as larger box sizes result in dimensions that are a similar size to the signal wavelength creating interfering box modes.\newline
\indent We have developed a Python model based on scikit-rf, an open source module to create a series of cascaded lumped-element networks connected by transmission lines to simulate an N-channel filterbank. Similar methods have already been developed to simulate and analyse FBSs \cite{ChePhdThesis2018} but only from the perspective of resonator parameters, rather than using physical properties as the simulation handles, such as geometry dimensions and dielectric thickness. The model produces similar microwave responses to Sonnet by extracting the microwave component and material properties from small scale Sonnet simulations to produce libraries for use across full, cascaded simulations. For example, the coupling capacitors were modelled as a lumped capacitance with parasitic series lumped inductances on either side with values calibrated against Sonnet simulations. A single filter typically taking  $\approx$10 minutes to simulate in Sonnet can be simulated in $<$ 5 seconds. The agreement between Sonnet simulations and the python model can be seen in Fig.~\ref{fig:SkrfModelComparisons}. This code enabled the tolerance analysis shown in Section~\ref{sec:FabErrorTolerance}, and allowed us to qualitatively observe the impact the fabrication errors have on the spectral response when errors are considered across multiple interacting (over-sampled) channels.\newline
\indent An example of the full FBS can be seen in Fig.~\ref{fig:filterbank120-180GHzFullFabError}. The impact of these fabrication errors is minor compared to that of  tan$\delta = 7\times 10^{-4}$, which accounts for much of the difference between the ideal and realistic FBS, displaying a large change in $\eta$ as expected from the results seen in Section~\ref{sec:LossTangentImpact}. Despite this, the FBS should still exhibit $\eta\approx$50\%.

\section{Conclusion}

We have presented a new filter design for a superconducting microstrip FBS and presented the optimisation process and the tolerance analysis of the design for a variety of key fabrication errors. The main differences in design compared to other FBSs \cite{SuperSpec2015, Deshima2019} include: the use of microstrip transmission line reducing loss by radiation loss, and parallel plate capacitor couplers at the end of each resonator rather than proximity coupling. Whilst also being a simple and easy to realise design, simulations show these filters have a high tolerance to fabrication error. The quality factor of each filter is most sensitive to variations in microstrip line widths (smallest width: 2$\mu$m) due to over or under etching, resulting in a worst case shift in $Q_\textrm{filt}$ of $\approx \pm 10\%$ for a pessimistic etch uncertainty of $\pm 0.2\,\mu\textrm{m}$. The resonant frequency of each filter is also suitably robust, with the largest shift in $f_0$ of $\approx \pm 3.5\%$ being due to a $\pm 0.02\,\textrm{pH/sq}$ deviation in the niobium kinetic inductance.

A python model, calibrated with Sonnet simulations was used to simulate the FBS and study the impact of fabrication errors on the spectral response. These simulations imply there will be limited impact on the FBS performance with a smoothly varying etch error and dielectric thickness over the wafer. Instead, improving the dielectric loss tangent is the aspect of the filter bank design that will yield the greatest improvements in optical efficiency. For example, the lossless FBS could achieve $\eta = 80\%$, whereas with the expected SiN tan$\delta$, we see this reduce to just under 50\% in the best case.

\begin{acknowledgements}
    This work was supported by the members of the SPT-SLIM collaboration and made use of scikit-rf, an open-source Python package for RF and Microwave applications.
    G. Robson acknowledges the continuing support from the Science and Technology Facilities Council Ph.D studentship programme under grant reference ST/S00033X/1.
    We thank the reviewers who have provided insightful comments leading to improvements to this paper.
    
   \noindent The datasets generated during and/or analysed during the current study are available from the corresponding author on reasonable request.

\end{acknowledgements}


\begin{thebibliography}{99}

\bibitem{SuperSpec2015}
S. Hailey-Dunsheath, E. Shirokoff, P. S. Barry, C. M. Bradford, G. Chattopadhyay, P. Day, S. Doyle, M. Hollister, A. Kovacs, H. G. LeDuc, P. Mauskopf, C. M. McKenney, R. Monroe, R. O'Brient, S. Padin, T. Reck, L. Swenson, C. E. Tucker, J. Zmuidzinas, \textit{Proc. SPIE} \textbf{9153} (2014). DOI: 10.1117/12.2057229.

\bibitem{Deshima2019}
A. Endo, K. Karatsu, Y. Tamura, T. Oshima, A. Taniguchi, T. Takekoshi, S. Asayama, T. J. L. C. Bakx, S. Bosma, J. Bueno, K. W. Chin, Y. Fujii, K. Fujita, R. Huiting, S. Ikarashi, T. Ishida, S. Ishii, R. Kawabe, T. M. Klapwijk, K. Kohno, A. Kouchi, N. Llombart, J. Maekawa, V. Murugesan, S. Nakatsubo, M. Naruse, K. Ohtawara, A. P. Laguna, J. Suzuki, K. Suzuki, D. J. Thoen, T. Tsukagoshi, T. Ueda, P. J. de Visser, P. P. van der Werf, S. J. C. Yates, Y. Yoshimura, O. Yurduseven, J. J. A. Baselmans, \textit{Nat Astron} \textbf{3}, 989–996 (2019). DOI: 10.1038/s41550-019-0850-8.

\bibitem{KarkareLTD2021}
K. S. Karkare, \textit{J. Low Temp. Phys} This Special Issue (2021).

\bibitem{BarryLTD2021}
P. S. Barry, \textit{J. Low Temp. Phys} This Special Issue (2021).

\bibitem{kov_cs_2012}
A. Kovács, P. S. Barry, C. M. Bradford, G. Chattopadhyay, P. Day, S. Doyle, S. Hailey-Dunsheath, M. Hollister, C. McKenney, H. G. LeDuc, N. Llombart, D. P. Marrone, P. Mauskopf, R. C. O'Brient, S. Padin, L. J. Swenson, and J. Zmuidzinas, \textit{Proc. SPIE} \textbf{8452} (2012). DOI: 10.1117/12.927160.

\bibitem{Shirokoff2012}
E. Shirokoff, P. S. Barry, C. M. Bradford, G. Chattopadhyay, P. Day, S. Doyle, S. Hailey-Dunsheath, M. I. Hollister, A. Kovács, C. McKenney, H. G. Leduc, N. Llombart, D. P. Marrone, P. Mauskopf, R. O'Brient, S. Padin, T. Reck, L. J. Swenson, J. Zmuidzinas, \textit{Proc. SPIE} \textbf{8452} (2012). DOI: 10.1117/12.927070.

\bibitem{Endo2013}
A. Endo, C. Sfiligoj, S. J. C. Yates, J. J. A. Baselmans, D. J. Thoen, S. M. H. Javadzadeh, P. P. van der Werf, A. M. Baryshev, and T. M. Klapwijk, \textit{Appl. Phys. Lett.} \textbf{103}, 032601 (2013). DOI: 10.1063/1.4813816.

\bibitem{DelftSpectrometer2019}
A. Endo, K. Karatsu, A. P. Laguna, B. Mirzaei, R. Huiting, D. Thoen, V. Murugesan, S. J. C. Yates, J. Bueno, N. V. Marrewijk, S. Bosma, O. Yurduseven, N. Llombart, J. Suzuki, M. Naruse, P. J. de Visser, P. P. van der Werf, T. M. Klapwijk, J. J. A. Baselmans, \textit{J. Astron. Telesc. Instrum. Syst.} \textbf{5} (3) 035004 (2019). DOI: 10.1117/1.JATIS.5.3.035004.

\bibitem{Sonnet}
Sonnet Suites, Version 17.56. Accessed 2021. Available: www.sonnetsoftware.com

\bibitem{Eom2012}
B. H. Eom, P. K. Day, H. G. LeDuc, and J. Zmuidzinas. \textit{Nat. Phys.} \textbf{8}, 623-627 (2012). DOI: 10.1038/nphys2356

\bibitem{GoaThesis2008}
J. Gao, 2008, PhD Thesis, Department of Physics, Caltech, Pasadena, CA, USA.

\bibitem{Pappas2011}
D. P. Pappas, M. R. Vissers, D. S. Wisbey, J. S. Kline, J. Gao, in IEEE Transactions on Applied Superconductivity, vol. \textbf{21}, no. 3, pp. 871-874, (2011), DOI: 10.1109/TASC.2010.2097578.

\bibitem{Martinis2005}
J. M. Martinis, K. B. Cooper, R. McDermott, M. Steffen, M. Ansmann, K. D. Osborn, K. Cicak, Se. Oh, D. P. Pappas, R. W. Simmonds, and C. C. Yu\
\textit{Phys. Rev. Lett.} \textbf{95}, 210503 (2005), DOI: 10.1103/PhysRevLett.95.210503.

\bibitem{PaikHanhee2010Rqdl}
H. Paika, K. D. Osborn, \textit{Appl. Phys. Lett.} \textbf{96}, 072505 (2010), DOI: 10.1063/1.3309703.

\bibitem{Hailey-DunsheathS2014OMoS}
S. Hailey-Dunsheath, P. S. Barry, C. M. Bradford, G. Chattopadhyay, P. Day, S. Doyle, M. Hollister, A. Kovacs, H. G. LeDuc, N. Llombart, P. Mauskopf, C. McKenney, R. Monroe, H. T. Nguyen, R. O’Brient, S. Padin, T. Reck, E. Shirokoff, L. Swenson, C. E. Tucker, J. Zmuidzinas, \textit{J. Low Temp. Phys.} \textbf{176}, 841–847 (2014). DOI: 10.1007/s10909-013-1068-2.

\bibitem{Laguna2021}
A. P. Laguna, K. Karatsu, D. Thoen, V. Murugesan, B. Buijtendorp, A. Endo, J. Baselmans, in IEEE Transactions on Terahertz Science and Technology, (2021) DOI: 10.1109/TTHZ.2021.3095429.

\bibitem{ChePhdThesis2018}
G. Che, 2018 PhD Thesis, Department of Physics, Arizona State University, Tucson, AZ, USA.





\end{thebibliography}
\end{document}